\documentclass[dvips]{acta}
\usepackage{supertabular,lscape,epsfig}
\usepackage{amssymb}
\usepackage{amsmath}

\SetPages{0}{0}
\SetVol{76}{2026}

\begin{document}

\begin{Titlepage}

\Title{A short-period binary OGLE-BLG-ELL-006503 showing slow variations in brightness}

\Author{S. M. Rucinski} {Department of Astronomy and Astrophysics, 
University of Toronto\\
50 St.\ George St., Toronto, ON M5S 3H4, Canada\\
e-mail: slavek.rucinski@utoronto.ca}

\Received{Month Day, Year}

\end{Titlepage}

\Abstract{Analysis of 28 years of  OGLE project data for the 0.404 day period binary 
OGLE-BLG-ELL-006503  (V1231~Sco) shows a superposition of tidal-interaction variations
of about 0.07 magnitude in the $I$-band onto a time-scale $5-10$ year, $0.15-0.17$ magnitude 
brightness drift.
Seasonal light-curves suggest that this long-term variability is driven by large photospheric 
spots ($\approx 0.3$ stellar radii) that exhibit a tendency to
orient themselves relative to the tidal interaction line. 
The eclipse time $O-C$ residuals reveal an orbital period change 
characterized by a timescale of approximately $2 \times 10^6$ 
years, suggesting a mild mass redistribution, potentially linked to global magnetic fields
reconfiguration. Gaia parallax and absolute magnitude 
estimates indicate a system consisting of solar-type stars.
The lack of eclipses and extreme spot activity suggest the binary 
may consist of late-type, tidally-deformed stars rather than a typical W~UMa-type system.}
{close binary stars, spotted stars, W~UMa binaries}

\section{Introduction}

Photospheric spots have long been suggested to exist on the surfaces of W~UMa-type binaries. 
Reports on their likely presence date back to the first sufficiently accurate, photoelectric 
observations (e.g.\ Brownlee 1957), 
where spots were invoked to explain small night-to-night changes in 
light curve shapes. At that time, observations required manual, real-time definition of the 
measured image and were largely irreproducible; mechanical errors in the telescope drive 
or pointing issues could be indicated as primary causes. Over time,
W~UMa-type binaries became frequent targets for investigations 
because they yield rapid results suitable for short-term thesis projects. 
While the resulting abundance of uncalibrated light curves 
hindered progress in studying intrinsic variations,  sufficient evidence accumulated 
to indicate that surface spots are a common feature of these systems.

Spot extent, location, and temporal evolution remained areas of speculation, often tied 
to  "light-curve solutions" following the development of the W~UMa binary
model by Lucy (1968a, 1968b). 
This model explained the uniformity of temperatures over a common envelope
despite different component masses and 
could accommodate the spots, albeit at the cost of 
increased parameter ambiguity. Consequently, 
information regarding the long-term evolution of  spots remained particularly scant.

The Optical Gravitational Lensing Experiment (OGLE) project (Udalski {\it et al,}  2015)
was recognized from its inception as a premier source for 
studies of stellar variability (Paczy\'{n}ski 1986). 
The current paper utilizes the high stability, uniformity, and temporal baseline of OGLE data 
to investigate the long-term effects of spots in the close binary OGLE-BLG-ELL-006503. 
The star was identified by Bohdan Paczy\'nski, the initiator of the project, 
through visual inspection of the first three seasons of the Phase-II of OGLE.
The star was noted as a short-period binary system exhibiting 
a progressive change in brightness (Ruci\'nski and Paczy\'nski 2002). 
Located in the the Baade's Window (BW), 
an area of relatively low interstellar extinction 
toward the Galactic bulge, the system showed a progressive 
brightening of over 0.1~mag in the near-infrared $I$ band, superimposed on binary 
short-period variations by about $0.07 - 0.08$~mag. 
The simplest explanation for this variability involved a reconfiguration 
of photospheric spots modulating the radiative flux of the binary. 
The original, tentative W~UMa-type classification was based on 
continuous variability with a relatively short orbital period of 0.40359 days. 

The current name of our object, OGLE-BLG-ELL-006503, 
refers to the primary database of eclipsing and ellipsoidal binary 
stars in the OGLE project (Soszy\'{n}ski {\it et al,} 2016). 
The star was identified in successive phases of the project as:
OGLE-II (1997-2000): BUL~SC27~23555, OGLE-III (2001-2009): BLG117.7.38113, 
OGLE-IV (since 2010): BLG604.09.56696
Although the binary is listed in the General Catalogue of Variable Stars as V1231~Sco,
here we continue to use the name OGLE-BLG-ELL-006503 as an homage to 
the OGLE project and its achievements.

\begin{figure}[t]
\centering
\includegraphics[scale=0.5]{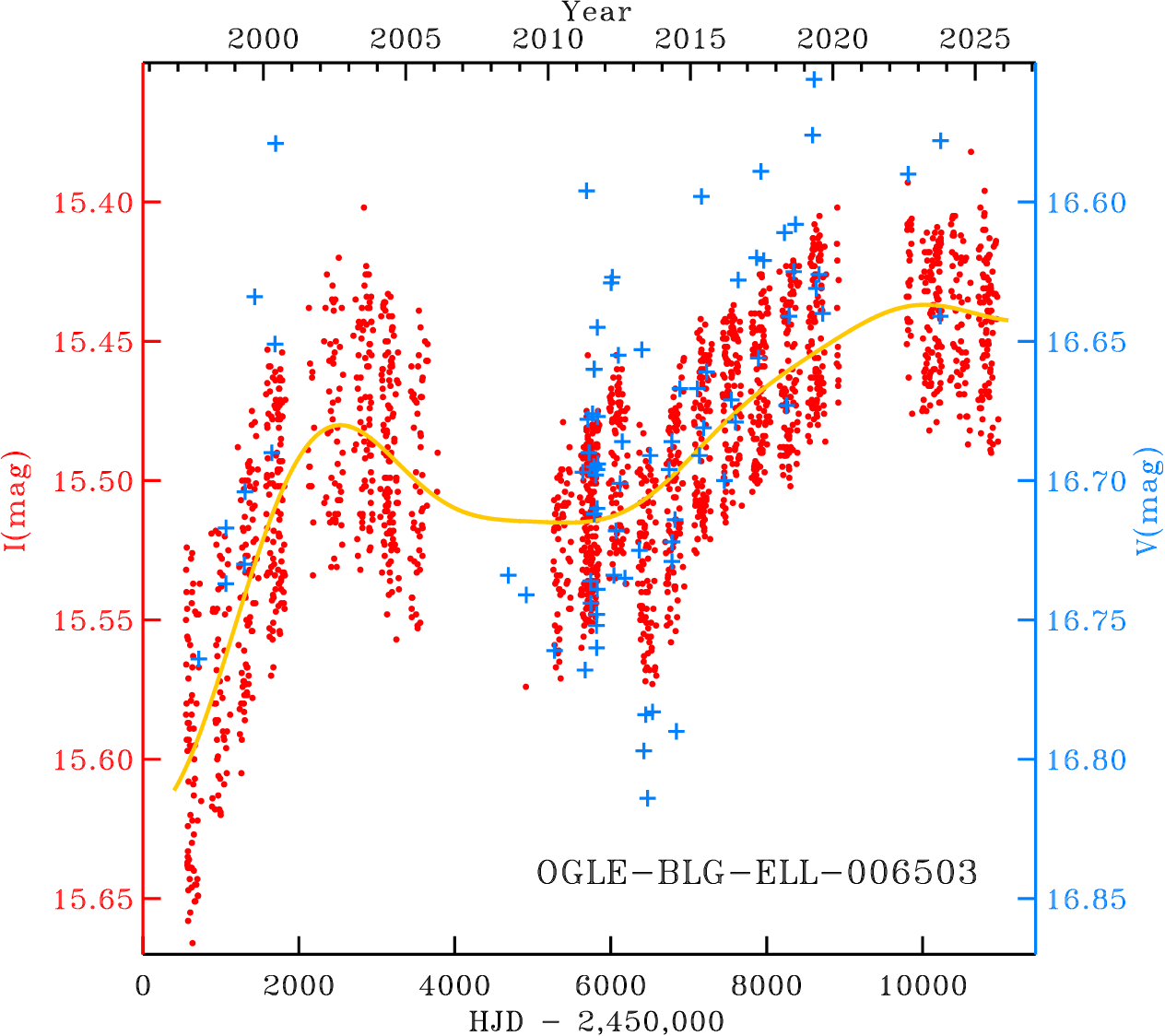}
\FigCap{The combined data for OGLE-BLG-ELL-006503 plotted versus 
time $T = HJD - 2\:450\:000$. 
The upper horizontal scale is expressed in years. 
The $I$-band data are small red points, while the $V$-band data are blue 
crosses. Note an offset of left and right vertical axes by +1.20 mag.}
\end{figure}

\section{The data}

The data used in this project consist of 1971 
observations in the $I$-band and 82 in the $V$-band, 
collected during 24 observing seasons in years 1997 to 2025. 
These data are available through the OGLE project 
webpage\footnote{https://ogledb.astrouw.edu.pl/~ogle/OCVS/getobj.php?s=OGLE-BLG-ELL-006503}.
The observed magnitudes in both photometric bands are shown in Fig.~1, 
plotted with an offset of 1.20 mag between the $I$ and $V$ bands which was
established by the analysis of the most abundant 2011 data (see below).
The data  exhibit orbital binary variability combined with a long-term 
trend. A smooth yellow line in the figure traces the median magnitude level calculated 
for available seasons over 28 years. 
While the median seasonal $I$-band magnitudes could be well 
established, the corresponding median levels for the $V$-band were poorly defined
due to large excursions of individual observations. possibly caused by
photospheric flares.

Formal errors for individual  observations in the database are quoted  at 
a median level of 0.006 in $I$ and 0.008 mag.\ in $V$, but 
some observations were more strongly affected by crowding in BW
during periods of poor seeing so that some errors reach up to
0.011 in $I$ and 0.013 in $V$. 
Systematic uncertainties in OGLE data are discussed in Skowron {\it et al.} (2016);
typically, they are the smallest -- at a level of 0.006 to 0.010 mag -- 
close to the magnitude range of our target.  

\begin{figure}[h]
\centering
\includegraphics[scale=0.38]{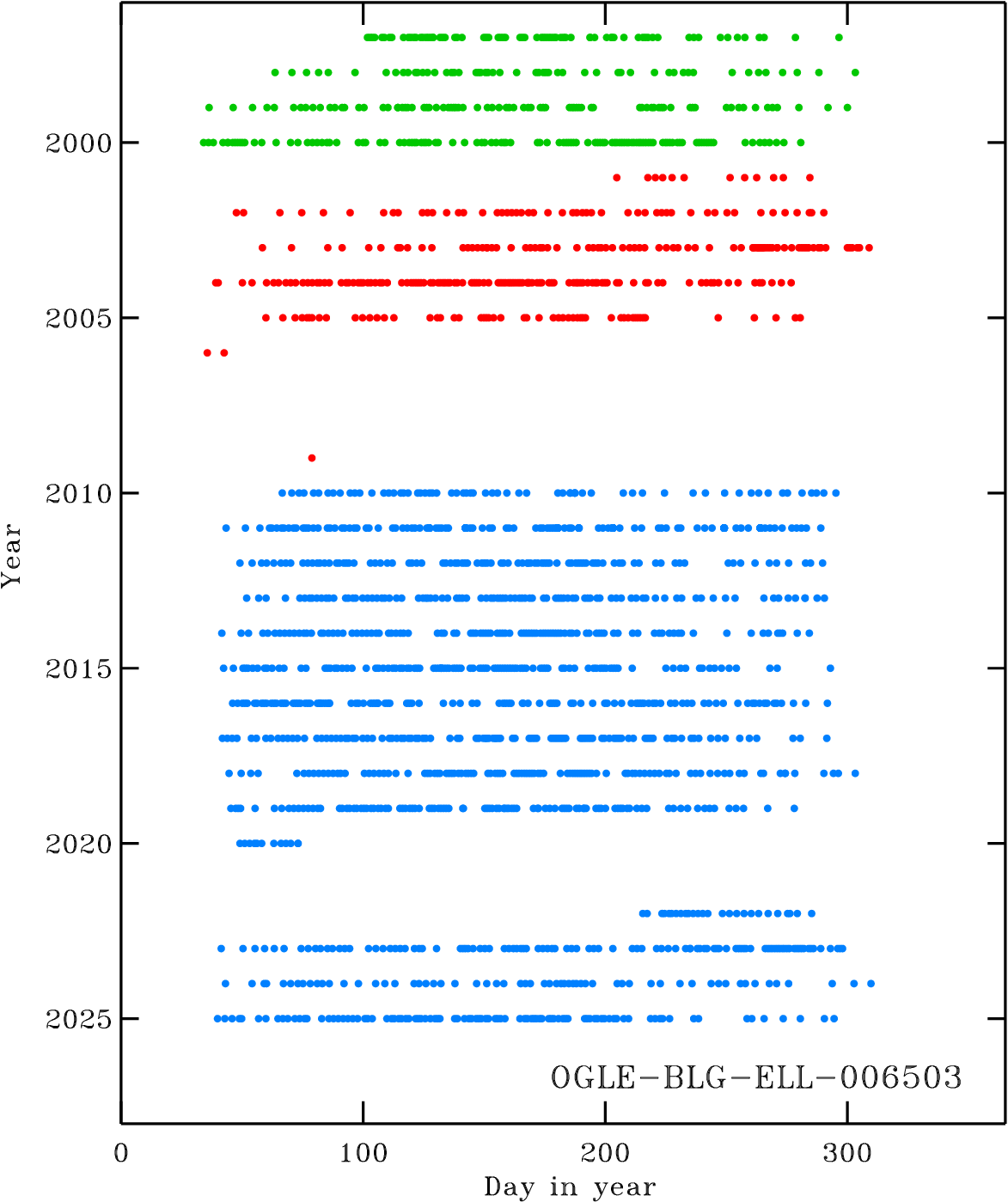}
\FigCap{The moments of the $I$-band observations are shown versus time (HJD - 2,450,000)
for each season to illustrate the density of the time coverage. 
The three colour codes correspond to the three phases of the OGLE project, 
II (green), III (red) and IV (blue). 
The previous paper (Ruci\'nski and Paczy\'nski 2002) used the data for the first
three of the phase II seasons.}
\end{figure}

The number of $I$-band observations per season (Fig.2) 
varied between 12 (in years 2001 
and 2020) and 198 (year 2011). With the two shortest runs removed, the 
shortest run consisted of 26 observations and the median number
of $I$-band observations per season was 95. 
No $V$-band observations were obtained on about half of the seasons and 
their number was always ten or more times smaller than in the $I$-band.

Long gaps took place in the years 2006 -- 2009 and 2020 -- 2022.
The former gap coincided with a relatively rapid and complex change in 
the combined system brightness around the switch from OGLE-III to OGLE-IV.
The latter gap was less detrimental as it apparently
corresponded to a period of slow evolution near the star light maximum; it
took place during the Covid-19 pandemic telescope shut down.
We note that the poor definition of $V$-band data and the high number of apparent 
outliers in Fig.1  may be caused not only by flare events, but possibly by
a larger amplitude of tidal variability for intrinsically red stars. 

Since the data were combined from different phases of the OGLE project, 
a legitimate question arises regarding the photometric consistency 
associated with the use of different cameras. 
Dr.\ Andrzej Udalski, the leader of the project, very kindly extracted 
the $V$ and $I$ data for eight stars in the field of our target 
that were assumed to be constant. 
Their constancy appears to be confirmed at the level of $0.01 - 0.02$ mag.

\section{The seasonal $I$-band light curves}

The seasonal light curves were sampled randomly and sparsely. 
The observations in the $I$-band were done 
with a median time interval of 1.97 day which
corresponds to almost five orbital periods of our target star. The $V$- and $I$-band 
data were collected independently so that temporal correlations could not be 
addressed.  

The season of 2011 with 198 $I$-band and 22 $V$-band observations
(Fig.3) was  exceptional in the density of $I$-band observations
(the median spacing of half hour) and with a relatively large number of
$V$-band observations (in other seasons the number was $<7$).
The 2011 data were used to estimate the observed mean colour index 
of the star as $V-I = 1.20 \pm 0.01$ mag. This was done by 
matching the main orbital variations in both spectral bands 
and assigning the infrequent (3 out of 22) $V$-band flux increases  to flares
(possibly affecting short wavelengths more strongly). 

\begin{figure}[t]
\centering
\includegraphics[scale=0.38]{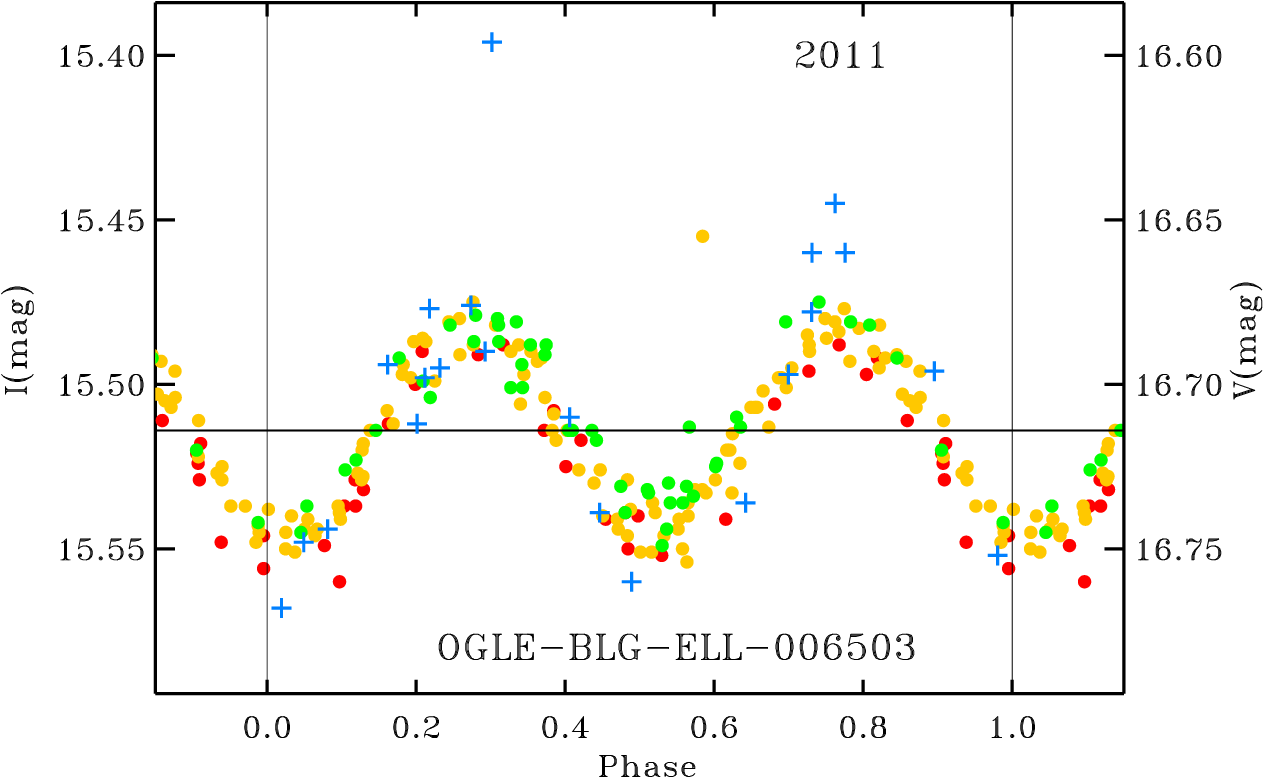}
\FigCap{The best observed seasonal $I$-band light curve 
of OGLE-BLG-ELL-006503 for year 2011
with the superimposed $V$ band observations (blue crosses) plotted with the offset of
$V-I = 1.20$ mag. To visualize possible temporal 
changes during the season, the $I$-band light curve has been divided into 
three equal-duration segments using colour symbols, red -- yellow -- green, 
from the earliest to the latest segment. 
}
\end{figure}

\begin{figure}[b]
\centering
\includegraphics[scale=0.35]{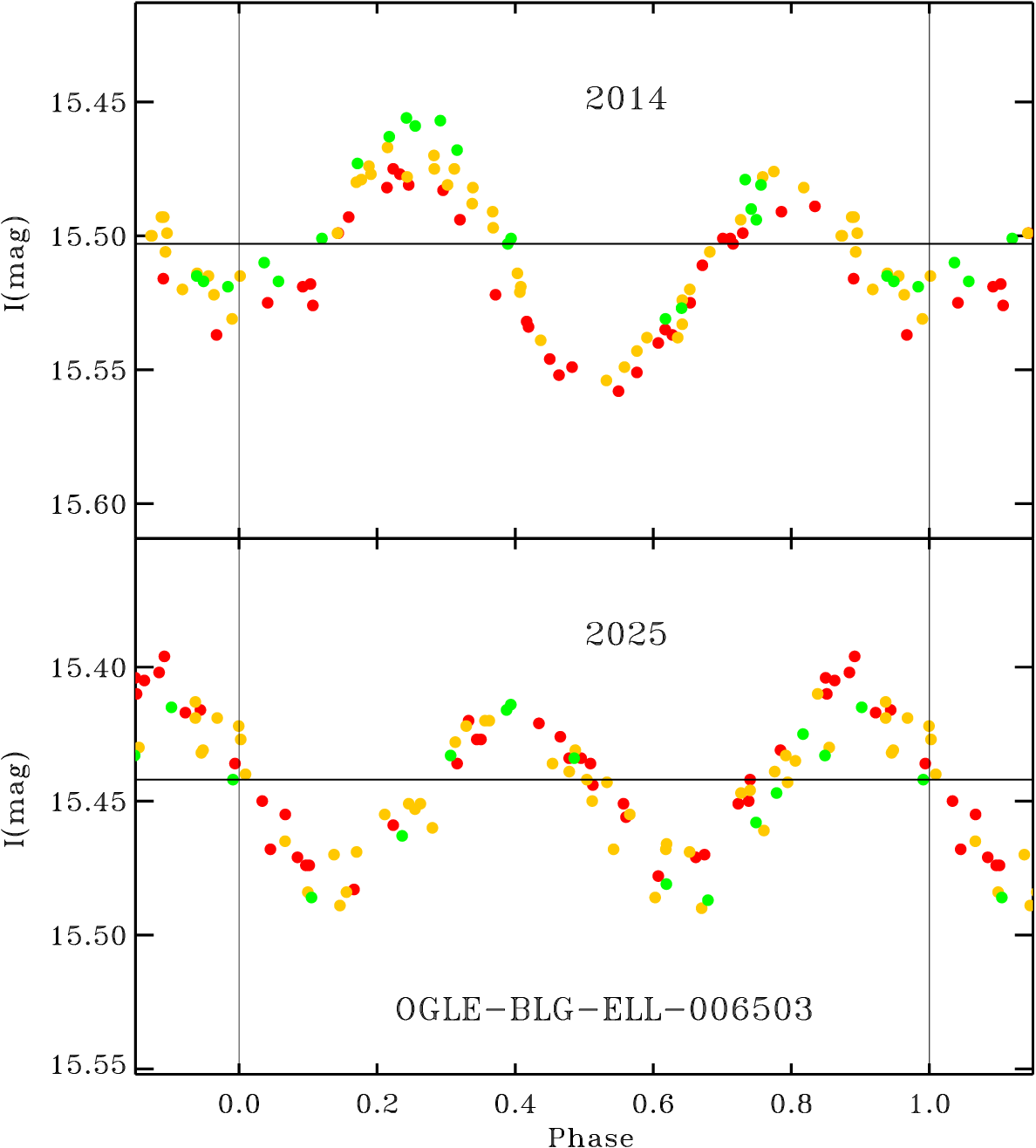}
\FigCap{$I$-band variability 
during the observing seasons of 2014 and 2025. 
The light curves have been divided into the  
equal-duration segments, as in Fig.3. 
Note that the normally deeper minimum was actually shallower 
of the two in 2014 and the phase shift by about 0.15 in 2025.}
\end{figure}

The seasonal $I$-band light curves over the 28 years of observations are
generally similar to that in 2011 (see Fig.3), but two seasons deserve
special attention.
The 2014 light curve underwent the strongest modification: the primary minimum, 
usually the deeper, became the shallower of the two. 
The curve from the last available 2025 season shows an obvious phase shift 
 by approximately 0.15 of the period. 
This shift appeared between 2010 and 2015 and continued in the following
years indicating a possible period lengthening; it is shown in Fig.5.

\begin{figure}[h]
\centering
\includegraphics[scale=0.35]{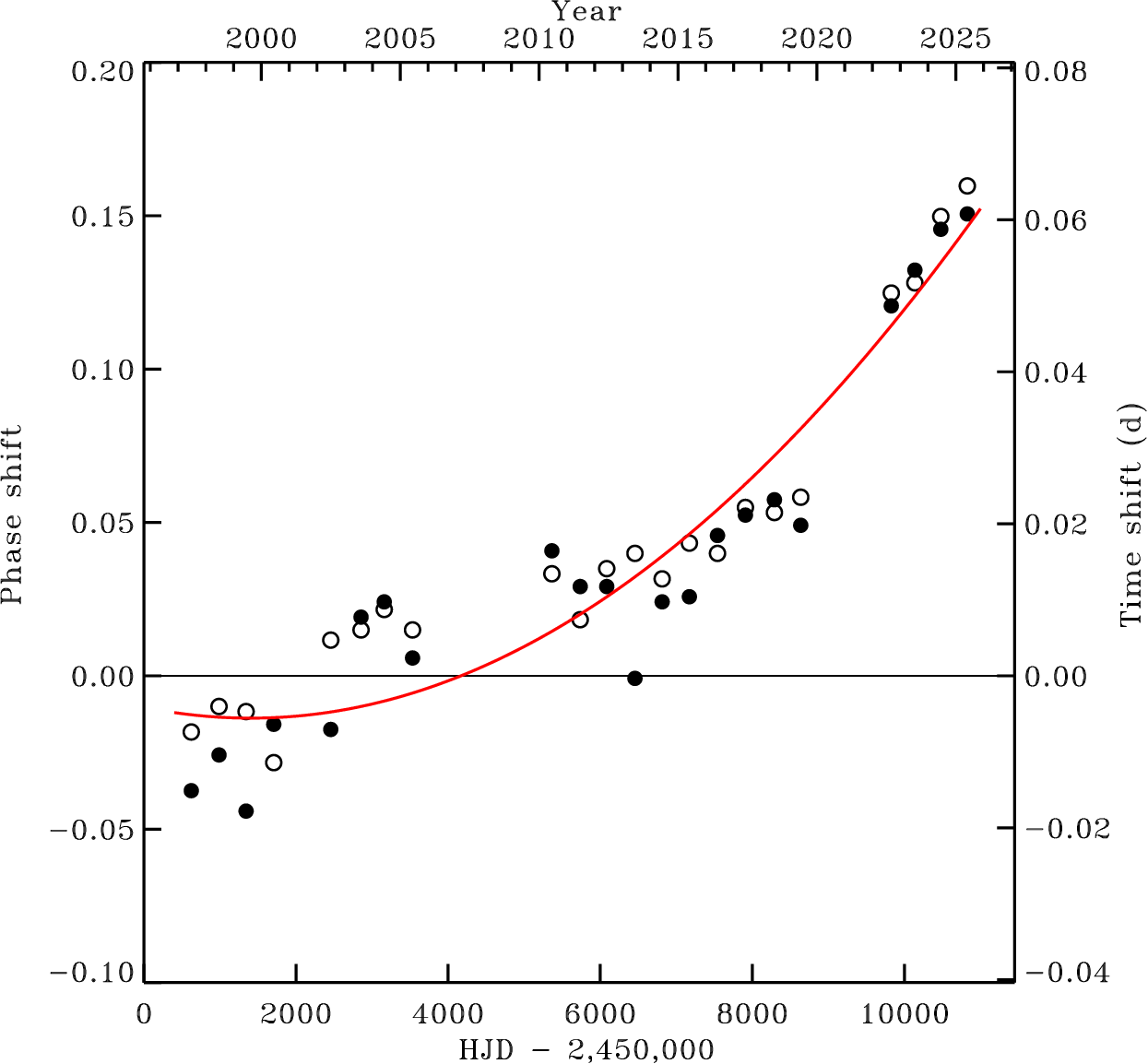}
\FigCap{The eclipse displacement from the linear ephemeris for both 
light minima, with the primary and secondary minima marked by filled 
and open circles respectively.}
\end{figure}

A quadratic term $a \times E^2$ has been added to the 
linear time-of-eclipse elements (Soszy\'nski {\it et al.}\ 2016), 
$T_0 - 2\,450\,000 = 7000.2029 + P \times E$, where 
$P = 0.4035877$ 
is the orbital period and $E$ is the number of orbital cycles. The fits 
of the phase shift vs.\ time, as in Fig.5, 
were subject to repeated bootstrap sampling solutions to estimate
the uncertainties. The corrections are:
$\Delta T_0 = -0.0041^{+0.0076}_{-0.0037}$ d,
$\Delta P = (-0.84^{+1.01}_{-1.07}) \times 10^{-6}$ d, and
$a = (+1.18^{+0.41}_{-0.36}) \times 10^{-10}$ d. 
The quadratic term implies the time scale of the systematic period change 
of $T \simeq 1/((dP/dE)/P) = (1.9 \pm 0.6) \times 10^6$ years. 
Such a scale is not unusual in solar-type close binaries, 
it may correspond to minor rearrangement of masses in the 
binary or to a global magnetic field reorganization.

\begin{figure}[h]
\centering
\includegraphics[scale=0.45]{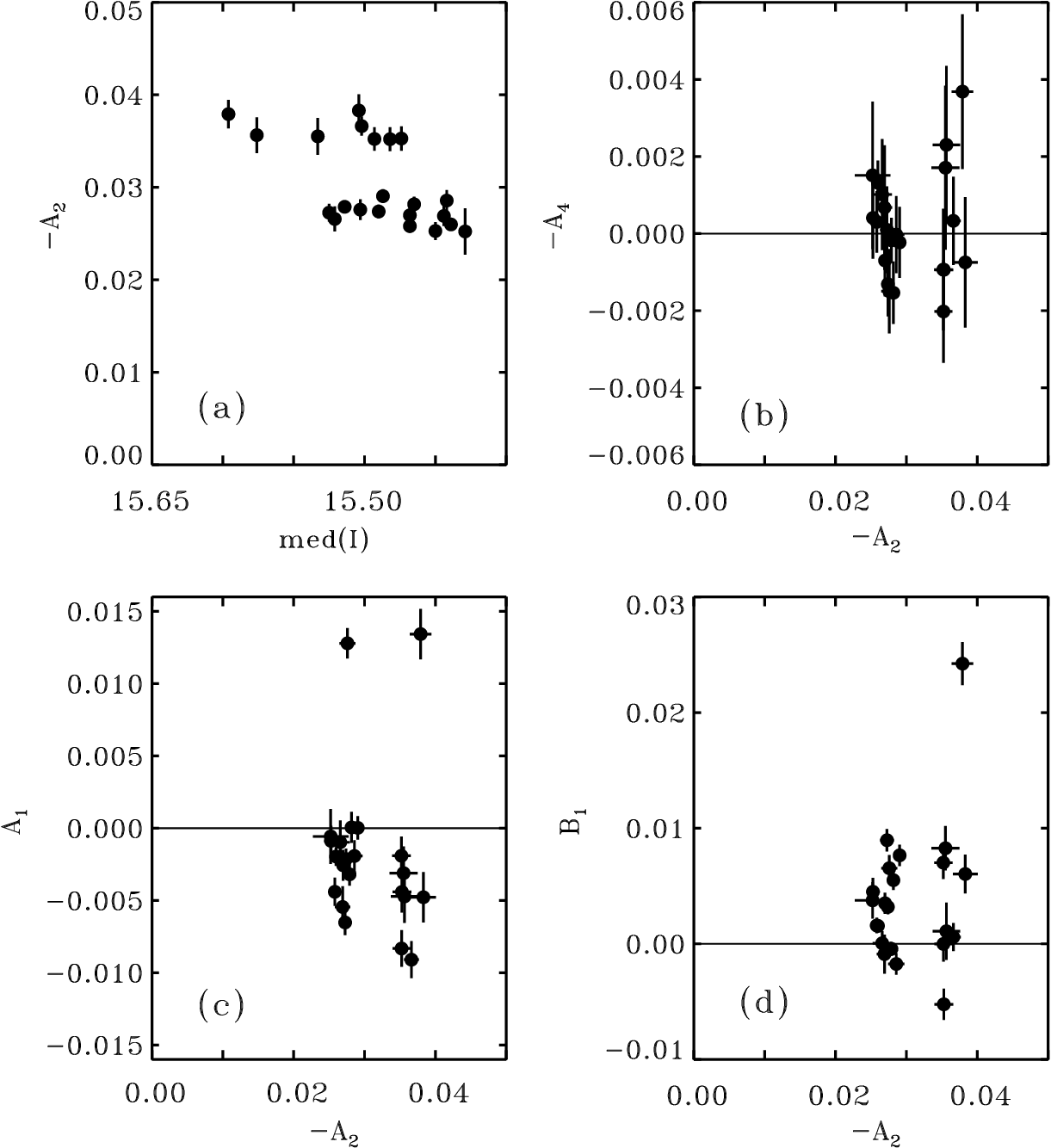}
\FigCap{Results of an approximation of the visible seasonal variability 
of the $I$-band flux
by a simplified trigonometric series (see the text). Panel (a) shows the coefficient
of the main, tidal-distortion variation ($\propto \cos 2\phi$) 
as a function of the seasonal median magnitude.  
The remaining panels show other terms as functions of $A_2$: In (b) the term $A_4$ 
representing sharpness of the light minima, in (c) $A_1$ reflecting the relative 
depth of minima, in (d) $B_1$ showing the main light-curve asymmetry.}
\end{figure}

Photospheric spots appear to drive the seasonal light curve changes and the 
mean brightness variations of OGLE-BLG-ELL-006503. 
With the discussed above phase shift 
accounted for, these changes can be characterized by a simplified 
expression similar to a Fourier series for the flux
variations (cf.\ Ruci\'{n}ski 1973, 1993). A reduced number of terms is appropriate in view
of the irregular sampling and the resulting loss of the data orthogonality:
$f(\phi)/f_{\rm max} = A_0 + A_1 \cos(\phi) + A_2 \cos(2\phi) + A_4 \cos(4\phi) + B_1 \sin(\phi)$.
The normalizing maximum flux, $f_{\rm max}$, was determined individually for each season.
Because of the rigidly set phase counting from the deeper minimum, the cosine
terms are mostly negative. The largest, $\cos(2\phi)$ term represents the 
tidal-distortion variability; thus, the $2 \times |A_2|$ can be considered 
as the dominant peak-to-peak flux variation. The  $\cos(\phi)$ term 
reflects relative depth of  minima, while the $\cos(4\phi)$ term may
sense a potential "down-spiking" of eclipses; the $\sin(\phi)$ term represents the global
light-curve asymmetry.

Results for the 22 available seasons are shown in Fig.6. In panel (a), the dominant term 
$A_2$ is plotted against seasonal median $I$ magnitudes. 
With the increased brightness (magnitudes decreasing to the right), 
the semi-amplitude $A_2$ was the smallest suggesting that spots may 
augment the tidal variability.  
When spots were present ($I > 15.47$ mag), data tended to group 
around two values of
$A_2$, $-0.027$ and $-0.037$, suggesting spots formed either 
along or perpendicular to the line joining the components. 
The light curves show no evidence of eclipses since $A_4$ remains near zero 
(panel {\it b} in Fig. 6). The primary minimum is generally the deeper one as originally
defined (panel {\it c} in Fig. 6); the exceptions 
were the 1997 and 2014 seasons (the latter shown
in Fig.4). Finally, $B_1$ (panel {\it d} in Fig. 6) was generally positive, 
which indicates that the first 
maximum (following the deeper minimum) was nearly always the brighter one. 
In interacting binaries, this is the side expected to receive matter 
from the expanding primary component due to the Coriolis deflection
(the O'Connell effect). 
The large positive $B_1$ outlier coincided with one of the large positive $A_1$ outliers.

\section{Properties of the binary}

The observed maximum brightness is estimated at $V=16.62$ mag and $I=15.42$ mag, 
$V-I=1.20$ mag (see Sec.3). 
The red colour is mostly due to interstellar reddening in the BW. 
Since the previous paper (Ruci\'nski and Paczy\'nski 2002),
 new tools have become available: (1)~the reddening/extinction calculator for 
 the BW (Nataf {\it et al.} 2013) and (2) the precise Gaia 
 parallaxes ({\it Gaia} Collaboration, Valleri {\it et al.} 2023). 
The parallax of $0.384 \pm 0.059$~mas  implies a distance of $2.60 \pm 0.40$~kpc. 
The OGLE extinction calculator predicts the extinction $A_I =0.829$ and reddening 
$E(V-I)=0.688$ to the bulge at the sky position of the star.
Assuming a linear increase to 4 kpc, we estimate the extinction
and reddening to the distance of OGLE-BLG-ELL-006503  at $A_I=0.54$ mag,
$A_V = 0.99$ mag and $E(V-I)=0.45$ mag.
The absolute magnitude estimates are done usually in the $V$-band.
The above numbers imply $M_V=+3.6 \pm 0.1$ mag
 and $(V-I)_0 =+0.75 \pm 0.05$ mag. This is consistent with two identical 
 solar-type stars and aligns with the period-luminosity relation for W~UMa binaries 
 (Mateo and Rucinski 2017), which predicts $M_V =+3.8 \pm 0.3$. 
 We note that for the Sun, $(V-I)_0 =+0.70 \pm 0.01$ mag so that the 
 binary may consist of slightly later-type stars than the Sun.

\section{Conclusions}

The star OGLE-BLG-ELL-006503 was selected for its systematic 
brightness changes, which was an unusual property for its presumed W~UMa type. 
This paper provides a firm confirmation of long-term variability by 15 to 17 percent 
over 28 years of the OGLE monitoring. This long-term trend couples with 
tidal orbital variability (about 5 to 7 percent) with a possibility that spots tend
to form predominantly along or at right angles to the tidal interaction line. 
The low amplitude and absence of eclipses suggest a binary seen at a
low orbital inclination.

It is currently not possible to confirm our initial assumption 
that OGLE-BLG-ELL-006503 is a W~UMa star (Rucinski and Paczynski 2002). 
Although small surface inhomogeneities are  observed in W~UMa binaries,
large spots such as on our target (up to about 0.3 radius)  are apparently
not common in these systems.
The David Dunlap Observatory spectroscopic program of 163 binaries
included 90 well-observed W~UMa-type binaries (Rucinski {\it et al.} 2013).
While we detected large spots in several short-period detached binaries, 
only one W~UMa system, V357~Peg,  showed large spots
in the radial-velocity profiles (Rucinski {\it et al.} 2008). 
The components of OGLE-BLG-ELL-006503 
appear to have equal effective temperatures -- which is a W UMa-type prerequisite -- 
a mass difference cannot be established in the
absence of eclipses and without a spectroscopic follow-up. 
Attractiveness of the star for further studies is considerably reduced by 
its difficult sky location for in-depth spectroscopic investigations.

The binary most likely belongs to the class of short-period RS~CVn stars, also
called BY~Dra type stars (for a recent statistics, see: Chahal {\it et al.}  (2022), which
are known to display strong spot and chromospheric activity. This is an inhomogeneous 
group of low-mass stars consisting of very close, detached binaries and single
stars rotating rapidly because of their youth. A possible bright analogue of 
OGLE-BLG-ELL-006503  would be then the binary
XY~UMa (eg.\ Pojma\'{n}ski, G. 1998). As pointed by St\c{e}pie\'{n} (2001), 
energy-carrying external circulation may lower or disrupt 
magnetic activity of W~UMa binaries.
Such a circulation is in fact the central feature of his model (St\c{e}pie\'{n} (2009)
which  appears to explain the complex  picture of typical W~UMa stars when analyzed
at high spectroscopic resolution (Rucinski 2025).

\Acknow{I would like to express my thanks to Drs. R.\ Poleski, I.\ Soszy\'{n}ski and
particularly A.\ Udalski for friendly, rapid and very helpful communications 
and extended advice on the use of the OGLE data.
Special thanks are due to Dr.\ K.\ St\c{e}pie\'{n} for his excellent comments and 
advice. The reviewer is to be thanked for very useful suggestions and corrections.
}


\end{document}